\documentclass[journal=jpclcd,manuscript=letter]{achemso}
\usepackage{chemformula}
\usepackage[T1]{fontenc} 
\usepackage{hyperref}
\usepackage{xcolor}
\usepackage{verbatim}
\usepackage{float}
\usepackage{mathrsfs}
\usepackage{amsmath}
\usepackage{amssymb}
\usepackage{graphics}
\usepackage{graphicx}
\usepackage{physics}

\newcommand{\CPTG}{
Chemical Physics Theory Group, Department of Chemistry,
and Center for Quantum Information and Quantum Control,
University of Toronto, Toronto, Ontario M5S 3H6, Canada}
\newcommand{\FICOMACO}{
Grupo de F\'{\i}sica Computacional en Materia Condensada, 
Escuela de F\'{\i}sica,  Facultad de Ciencias, 
Universidad Industrial de Santander, Cra 27 calle 9, Bucaramanga, Colombia.}

\author{Leonardo F. Calder\'on}
\email{leonardo.calderon@utoronto.ca}
\affiliation{\CPTG}
\alsoaffiliation{\FICOMACO}
\author{Chern Chuang}
\email{chern.chuang@utoronto.ca}
\affiliation{\CPTG}
\author{Paul Brumer}
\email{paul.brumer@utoronto.ca}
\affiliation{\CPTG}
\title[Electronic-vibrational resonance in the NESS]
{Electronic-vibrational resonance does not alter steady-state
transport in natural light-harvesting systems}
\date{\today}

\begin{document}

\begin{tocentry}
\includegraphics[scale=0.25]{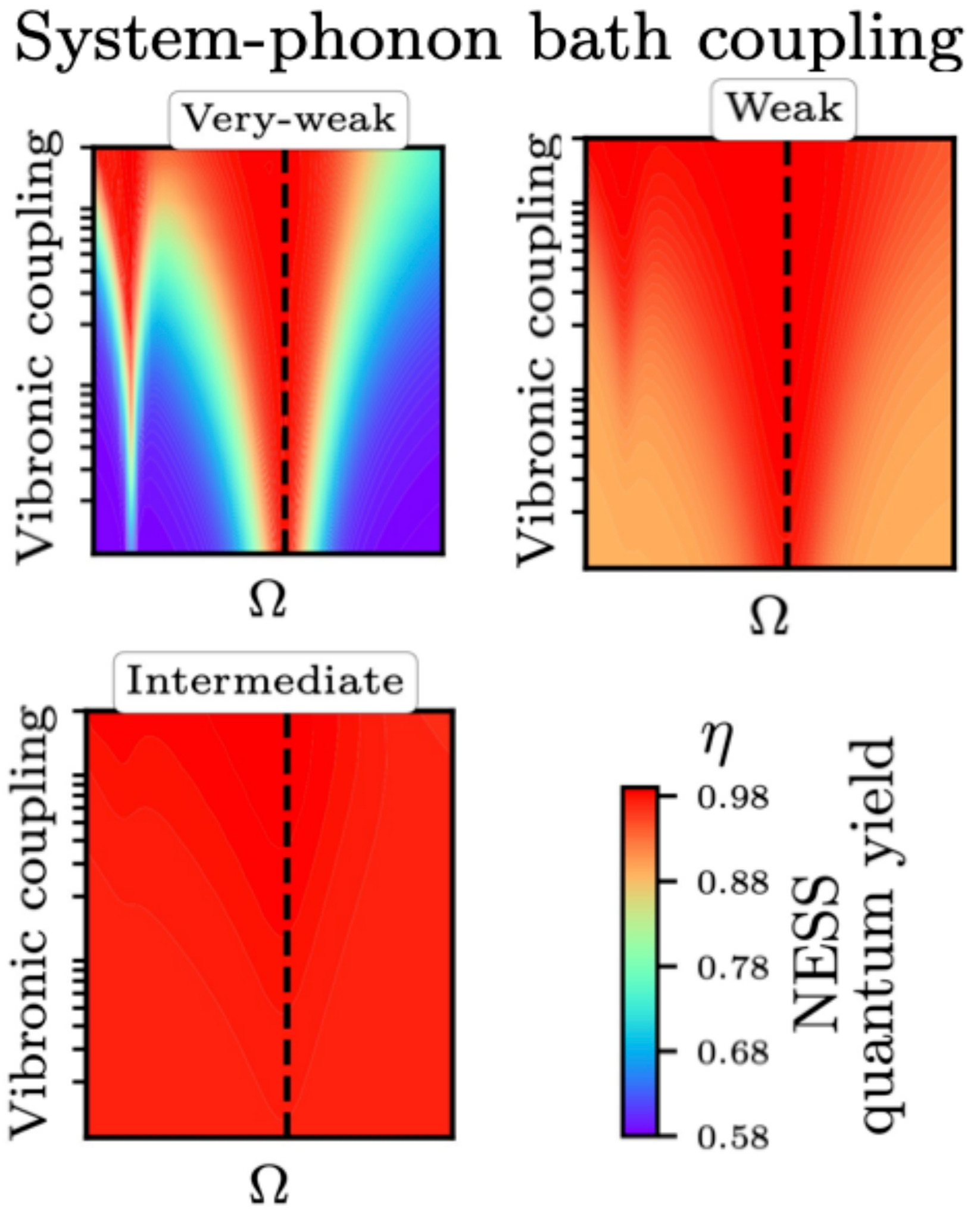}
\end{tocentry}

\begin{abstract}
Oscillations in time-dependent 2D electronic spectra appear as evidence
of quantum coherence in light-harvesting systems related to 
electronic-vibrational resonant interactions. 
Nature, however, takes place in a non-equilibrium steady-state, so the 
relevance of these arguments to the natural process is unclear. 
Here we examine the role of intramolecular vibrations in the non-equilibrium 
steady-state of photosynthetic dimers in the natural scenario of incoherent 
light excitation. 
It is found that vibrations resonant with the energy 
difference between exciton states do not increase the quantum yield nor the 
imaginary part of the intersite coherence that is relevant for transport 
compared with non-resonant vibrations in the natural non-equilibrium steady 
state.
That is, the vibration-electronic resonance interaction does not alter 
energy transport under natural incoherent-light excitation conditions.
\end{abstract}

Ultrafast spectroscopy has provided crucial information about the 
specific vibronic structure of biological light-harvesting systems and 
their interaction with the surrounding environment 
\cite{RMK01,Jon03,BS&05,CW&10,JM18,Man20}. 
The role of intramolecular vibrations in energy transfer has been examined 
in coherent pulsed laser excitation studies, where it has been argued that 
the electronic-vibrational resonant interaction 
can be responsible for energy transfer enhancements, the character 
of quantum coherence associated with nonlinear spectroscopic signals,  
and for non-trivial quantum behaviour of collective pigment motions
\cite{WM11,BVA12,CK&12,KO&12,CC&13,TPJ13,CP&13,FP&14,BVA14,HJ&14,OO14,RA&14,
DW&15,NN&15,SI&15,MS&16,DM&16,YH&19,CC&20,AY&20,HL&21,PN&22}. 
However, these laboratory-designed/controlled conditions differ fundamentally 
from nature \cite{JB91,MV10,BS12,PB12,KYR13,PBB17,Bru18,DB21}, where the 
light-harvesting systems are continuously illuminated
by incoherent natural radiation (sunlight), leading
to a non-equilibrium steady-state (NESS) \cite{TB18,Bru18,DB21,CB20,JB20}.
The key issue addressed in this letter is the extent to which
vibration-electronic resonance affects electronic energy transport
in the NESS.

We consider a prototype photosynthetic dimer to evaluate the impact
of the electronic-vibrational resonance and vibronic coupling under
natural incoherent light excitation. 
Specifically, we analyze a model PEB dimer in the cryptophyte algae PE545 
antenna protein \cite{DM&04,ND&10,Bla14}.
Each chromophore ($\mathrm{PEB_{50/61C}}$ and $\mathrm{PEB_{50/61D}}$)
is modelled as a two-level system coupled to a quantized high-energy
intramolecular vibrational mode and is excitonically coupled with one another. 
We solve for the steady-state of this vibronic dimer (system) using an  
open quantum system approach. 
We model the sunlight as a blackbody radiation bath, the protein/solvent 
environment as a thermal phonon bath, and consider other processes such as 
exciton recombination and exciton harvesting (trapping at the 
reaction center), acting on nanosecond and 
picosecond time-scales, respectively, via effective Lindbladians 
\cite{MKT14,TB18,JB20,CB20,JM20a}.

In this Letter, we analyze the different non-unitary contributions
from the environment on the NESS and its relation
to the vibronic effects. 
First, the electronic-vibrational resonance pattern is analyzed by computing 
the quantum yield for different Huang-Rhys factors, 
intramolecular vibrational frequencies, as well as the system-phonon bath coupling strengths. 
The quantum yield is contrasted with the imaginary part of the intersite
coherence, which dictates the transport between the chromophores.
We then compute the dynamics assuming a coherent excitation initial 
condition to look for similarities in the vibronic 
resonance pattern concerning steady-state results with  
different system-phonon bath coupling strengths. 
In doing so, we demonstrate that 
\emph{the vibration-electronic resonance interaction does not 
alter the NESS energy transport under natural incoherent-light
excitation conditions.}

\textbf{\textit{Model}}.\textemdash
Light-induced energy transport in molecular aggregates depends on the 
interaction between the electronic and nuclear degrees of freedom, the latter 
being associated with intermolecular and intramolecular vibrations 
\cite{MK11,BVA14}.
As is done here, the intermolecular vibrations are usually treated as
a low-frequency continuous phonon bath described by a spectral density,
and the intramolecular vibrations as discrete underdamped high-frequency
vibrations \cite{BVA12,BVA14}. 
In what follows, we denote the coupling and resonance interaction between
electronic excitations and intramolecular vibrations as vibronic coupling and vibronic resonance, respectively.

Consider a prototype photosynthetic vibronic dimer system, immersed within
a protein-solvent environment and excited by incoherent sunlight. 
For the $i^{\mathrm{th}}$ chromophore (site/molecule) only the electronic 
ground state $|g_{i}\rangle$ and the electronic first excited state 
$|\varepsilon_{i}\rangle$ are taken into account (two-level system 
approximation), and a single intramolecular vibrational mode of frequency 
$\Omega_{i}$ per chromophore is considered. 
The electronic creation (annihilation) operators
$\hat{\varphi}_{i}^{+}\left(\hat{\varphi}_{i}^{-}\right)$  are defined by
$|\varepsilon_{i}\rangle = 
\hat{\varphi}_{i}^{+}|g_{i}\rangle,\,|g_{i}\rangle =
\hat{\varphi}_{i}^{-}|\varepsilon_{i}\rangle$.
The intramolecular vibrational creation (annihilation) operators
$\hat{c}_{i}^{\dagger}\left(\hat{c}_{i}\right)$ then define the 
intramolecular vibrational excited state $|\chi_{i,\nu}\rangle=
\left(\hat{c}_{i}^{\dagger}\right)^{\nu}/\sqrt{\nu!}|0_{i}\rangle$, 
where $\nu$ is the vibrational quantum number, and $|0_{i}\rangle$ stands
for the vibrational ground state. 
The Hamiltonian of the $i^{\mathrm{th}}$ chromophore reads 
\begin{equation}
\hat{H}_{i} =
\varepsilon_{i}\hat{\varphi}_{i}^{+}\hat{\varphi}_{i}^{-}
+ \hbar\mathcal{G}_{i}\hat{\varphi}_{i}^{+}
  \hat{\varphi}_{i}^{-}(\hat{c}_{i}^{\dagger}+\hat{c}_{i})
+\hbar{}_{i}\Omega_{i}\hat{c}_{i}^{\dagger}\hat{c}_{i},
\end{equation}
where $\mathcal{G}_{i}=\Omega_{i}\sqrt{S_{i}}$ is the vibronic
coupling at frequency $\Omega_{i}$, and $S_{i}$ is the dimensionless 
Huang-Rhys factor that characterizes the electron-vibration coupling.

The global system-bath Hamiltonian reads 
$\hat{H}=\hat{H}_{\mathrm{Sys}}+\hat{H}_{\mathrm{Sys-B}}+\hat{H}_{\mathrm{B}}$.
Consider then a vibronic dimer, which is the system of interest from an open 
quantum system perspective, composed of two interacting chromophores
$A$ (acceptor) and $D$ (donor), where the intersite interaction
is assumed to be dipolar and represented by the electronic coupling
$V_{AD}$ ($V_{AA}=V_{DD}=0$).
The vibronic dimer Hamiltonian is given by
\begin{equation}
\hat{H}_{\mathrm{Sys}} =
\sum_{i}^{A,D}\varepsilon_{i}\hat{\varphi}_{i}^{+}\hat{\varphi}_{i}^{-}
+\sum_{i\neq j}^{A,D}V_{ij}\hat{\varphi}_{i}^{+}\hat{\varphi}_{j}^{-}
+\sum_{i}^{A,D}\hbar\mathcal{G}_{i}\hat{\varphi}_{i}^{+}
 \hat{\varphi}_{i}^{-}(\hat{c}_{i}^{\dagger}+\hat{c}_{i})
+\sum_{i}^{A,D}\hbar\Omega_{i}\hat{c}_{i}^{\dagger}\hat{c}_{i}.
\label{eq:Hs}
\end{equation}
The intramolecular vibrational degrees of freedom are treated
explicitly in the vibronic dimer model \cite{BVA12,BVA14}, while we discuss an electronic dimer model without the former in the Supporting Information.
Several discrete intramolecular vibrations are reported for 
cryptophyte algae \cite{DM&04,ND&10}.
Here, we performed the simulations considering different frequencies and 
Huang-Rhys factors to get an overall picture of the impact of the 
intramolecular vibrations on the NESS under incoherent light 
illumination. 
The remaining intermolecular vibrations are considered as parts of a thermal phonon 
bath described by a spectral density (see below).

We consider a model of the two phycoerythrobilin $\mathrm{PEB_{50/61C}}$ 
(acceptor) and $\mathrm{PEB_{50/61D}}$ (donor) chromophores (PEB dimer) from 
the protein-antenna phycoerythrin 545 (PE545) in cryptophyte algae. 
There is a large energy gap between the monomer electronic excited states
$\Delta E_{\varepsilon}=1042\,\mathrm{cm}^{-1}$ and a small excitonic 
coupling $V_{AD}=92\,\mathrm{cm}^{-1}$ between the PEB dimer. 
Exciton states are highly localized over a specific chromophore, and 
the exciton energy gap is $\Delta E_{e}=1058\,\mathrm{cm}^{-1}$. 
The transition dipole moments are $12.17\,\mathrm{D}$ ($\mathrm{PEB_{50/61C}}$) 
and $11.87\,\mathrm{D}$ ($\mathrm{PEB_{50/61D}}$). 
We assume that both chromophores have the same intramolecular vibrational 
frequencies and vibronic couplings, i.e., 
$\Omega=\Omega_{A}=\Omega_{D}$
and $\mathcal{G}=\mathcal{G}_{A}=\mathcal{G}_{D}$. 
The first three levels of quantized intramolecular vibrations are considered 
in the simulations below.

\textit{Interaction with the environment.}\textemdash 
The system-bath and bath Hamiltonians read 
\begin{equation}
\label{eq:Hsb}
\hat{H}_{\mathrm{Sys-B}} = 
\sum_{i}^{A,D}\sum_{l}\hbar g_{il}^{(\mathrm{e})}
\hat{\varphi}_{i}^{+}\hat{\varphi}_{i}^{-}
\left(\hat{b}_{l}^{(i)}+{\hat{b}_{l}}^{(i)\dag}\right)
+\sum_{i}^{A,D}\sum_{m}\hbar g_{im}^{(\mathrm{v})}
(\hat{c}_{i}^{\dagger}+\hat{c}_{i})
\left(\hat{b}_{m}^{(i)}+{\hat{b}_{m}}^{(i)\dag}\right)
-\sum_{i}^{A,D}\hat{\boldsymbol{\mu}}_{i}\cdot\hat{\mathbf{E}}(t),
\end{equation}
\begin{equation}
\label{eq:Hb}
\hat{H}_{\mathrm{B}}=  
\sum_{i}^{A,D}\sum_{l}\hbar\omega_{l}^{(i)}
    \hat{b}_{l}^{(i)\dag}\hat{b}_{l}^{(i)}
+\sum_{i}^{A,D}\sum_{m}\hbar\omega_{m}^{(i)}
    \hat{b}_{m}^{(i)\dag}\hat{b}_{m}^{(i)}
+\sum_{\mathbf{k},s}\hbar ck\hat{a}_{\mathbf{k},s}^{\dag}\hat{a}_{\mathbf{k},s}.
\end{equation}
Here, $g_{il}^{(\mathrm{e})}$ ($g_{im}^{(\mathrm{v})}$) represents
the coupling between the electronic (intramolecular vibrational) degrees
of freedom of the $i^{\mathrm{th}}$ chromophore and the $l^{\mathrm{th}}$
($m^{\mathrm{th}}$) phonon mode, where 
$\hat{b}_{l}^{(i)\dag},\hat{b}_{m}^{(i)\dag}$
$\big(\hat{b}_{l}^{(i)},\hat{b}_{m}^{(i)}\big)$ are the creation
(annihilation) operators of the phonon modes of frequencies 
$\omega_{l}^{(i)},\omega_{m}^{(i)}$
that couple to the electronic and intramolecular vibrational degrees
of freedom of the $i^{\mathrm{th}}$ chromophore. 
The electric field of the radiation \cite{MW95} is given by 
$\hat{\mathbf{E}}(t)=\hat{\mathbf{E}}^{(+)}(t)+\hat{\mathbf{E}}^{(-)}(t)$, 
with $\hat{\mathbf{E}}^{(+)}(t)= 
\mathrm{i}\sum_{\mathbf{k},s}
\left(\frac{\hbar\omega}{2\epsilon_{0}V}\right)^{1/2}\hat{a}_{\mathbf{k},s}
(\varepsilon_{\mathbf{k},s})\mathrm{e}^{-\mathrm{i}\omega t}$
and $\hat{\mathbf{E}}^{(-)}(t)=\left[\hat{\mathbf{E}}^{(+)}(t)\right]^{\dagger}$,
with $\hat{a}_{\mathbf{k},s}^{\dag}$ ($\hat{a}_{\mathbf{k},s}$)
being the creation (annihilation) operator for the $\mathbf{k}^{\mathrm{th}}$
radiation field mode in the $s^{\mathrm{th}}$ polarization state,
and $\hat{\boldsymbol{\mu}}_{i}$ is the dipole operator for the 
$i^{\mathrm{th}}$ chromophore.

The open quantum system dynamics of the density operator $\hat{\rho}$ for the 
vibronic dimer system described in Eq.~\ref{eq:Hs} is given by the master 
equation 
\begin{equation}
\label{eq:Mastereq}
\frac{\partial}{\partial t}\hat{\rho} =
-\frac{\mathrm{i}}{\hbar}\big[\hat{H}_{\mathrm{Sys}},\hat{\rho}\big]
+\mathscr{L}_{\mathrm{RB}}\left[\hat{\rho}\right]
+\mathscr{L}_{\mathrm{PB}}\left[\hat{\rho}\right]
+\mathscr{L}_{\mathrm{rec}}\left[\hat{\rho}\right]
+\mathscr{L}_{\mathrm{trap}}\left[\hat{\rho}\right].
\end{equation}
The first term on the right-hand side describes the unitary evolution
of the dimer system. 
The Liouvillian operator for the incoherent radiation
environment (blackbody radiation bath denoted as RB) and the protein/solvent
environment (phonon baths denoted as PB) read 
$\left(\mathscr{L}_{\mathrm{\mathrm{RB,PB}}}
\left[\hat{\rho}\right]\right)_{ij}=
-\sum_{kl}R_{ij,kl}^{\mathrm{RB,PB}}\rho_{kl}(t)$,
where we use the standard non-secular Redfield approach 
for thermal baths comprised of harmonic modes \cite{Nit06,MK11}
\begin{equation}
R_{ij,kl}^{\mathrm{\mathrm{RB,PB}}}=
\delta_{ik}\sum_{m}\Gamma_{jm,ml}^{\mathrm{\mathrm{RB,PB}}}(\omega_{lm})
+\delta_{jl}\sum_{m}\Gamma_{im,mk}^{\mathrm{\mathrm{RB,PB}}}(\omega_{km})
-\Gamma_{ki,jl}^{\mathrm{\mathrm{RB,PB}}}(\omega_{lj})
-\Gamma_{lj,ik}^{\mathrm{\mathrm{RB,PB}}}(\omega_{ki}).
\end{equation}

The damping matrix elements 
\begin{equation}
\Gamma_{ij,kl}^{\mathrm{RB,PB}}(\omega)=
\sum_{u,v}\int_{0}^{\infty}\mathrm{d}t\mathrm{e}^{\mathrm{i}\omega t}
C_{u,v}^{\mathrm{RB,PB}}(t)K_{u,ij}^{\mathrm{RB,PB}}K_{v,kl}^{\mathrm{RB,PB}},
\end{equation}
determine the time span for correlations. 
Here, $K_{u,ij}^{\mathrm{RB,PB}}$ denote the observables of the vibronic 
dimer system that are coupled to the radiation and phonon baths 
modes (see Eq.~\ref{eq:Hsb}), and 
\begin{equation}
C_{u,v}^{\mathrm{RB,PB}}(t)=
\int_{0}^{\infty}\mathrm{d}\omega\,
\omega^{2}J_{i}^{\mathrm{\mathrm{RB,PB}}}(\omega)
\bigg[\coth\left(\frac{\hbar\omega\beta}{2}\right)
\cos(\omega t)-\mathrm{i}\sin(\omega t)\bigg],
\end{equation}
the correlation functions of the radiation and phonon baths. 
Formally, the radiation bath is described by a super-Ohmic spectral density
with cubic-frequency dependence which generates long-lasting coherent
dynamics provided by the lack of pure dephasing dynamics and the strong
dependence of the decoherence rate on the system level spacing \cite{PBB17}
\begin{equation}
\label{eq:spectral_density_RB}
\omega^{2}J_{j}^{\mathrm{\mathrm{RB}}}(\omega)=
\frac{2\hbar\omega^{3}}{3(4\epsilon_{0}\pi^{2}c^{3})}.
\end{equation}

The phonon baths introduce dissipation and decoherence into the electronic
$(\mathrm{e})$ and intramolecular vibrational $(\mathrm{v})$ degrees
of freedom of the system Hamiltonian (Eq.~\ref{eq:Hs}). 
Its effects are encoded in the Drude-Lorentz spectral density 
\begin{equation}
\label{eq:spectral_density_PB}
\omega^{2}J_{j}^{\mathrm{\mathrm{PB}}}(\omega)=
\frac{2\gamma_{j}^{(\mathrm{e},\mathrm{v})}
    \lambda_{j}^{(\mathrm{e},\mathrm{v})}\omega}
{\hbar(\omega^{2}+\gamma_{j}^{(\mathrm{e},\mathrm{v})2})},
\end{equation}
where $\gamma^{(\mathrm{e},\mathrm{v})}$ represents the cutoff frequency
and $\lambda^{(\mathrm{e},\mathrm{v})}$ the reorganization energies.
The same spectral densities are assumed on each chromophore, with
$\gamma_{A,D}^{(\mathrm{e},\mathrm{v})}=100\,\mathrm{cm}^{-1}$ and different
values for $\lambda_{A,D}^{(\mathrm{e},\mathrm{v})}$ below. 
The temperature assumed for the phonon bath it is 
$T^{\mathrm{PB}}=300\,\mathrm{K}$,
and $T^{\mathrm{RB}}=5600\,\mathrm{K}$ for the blackbody radiation bath. 
The transition dipole moments of the chromophores are considered 
constants in time and parallel to the incoherent light electric field.

The exciton recombination and the trapping at the reaction center 
processes are considered localized processes on each chromophore 
\cite{MKT14,TB18,JB20,CB20,JM20a} and modelled through the effective 
Lindbladians 
\begin{equation}
\label{eq:Lrecloc}
\mathscr{L}_{\mathrm{\mathrm{rec}}}\left[\hat{\rho}\right]=
\tau_{\mathrm{rec}}^{-1}\sum_{i}^{A,D}
\left(|g_{i}\rangle\langle\varepsilon_{i}|\hat{\rho}|\varepsilon_{i}
\rangle\langle g_{i}|-\frac{1}{2}\left[|\varepsilon_{i}\rangle\langle
\varepsilon_{i}|,\hat{\rho}\right]_{+}\right),
\end{equation}
\begin{equation}
\label{eq:LRCloc}
\mathscr{L}_{\mathrm{\mathrm{trap}}}\left[\hat{\rho}\right]=
\tau_{\mathrm{trap}}^{-1}\left(|\mathrm{RC}\rangle\langle
\varepsilon_{A}|\hat{\rho}|\varepsilon_{A}\rangle\langle\mathrm{RC}|
-\frac{1}{2}\left[|\varepsilon_{A}\rangle\langle
\varepsilon_{A}|,\hat{\rho}\right]_{+}\right),
\end{equation}
where $\left[\mathcal{\hat{O}}_{1},\mathcal{\hat{O}}_{2}\right]_{+}$
represents the anticommutator between operators $\mathcal{\hat{O}}_{1}$
and $\mathcal{\hat{O}}_{2}$. 
The recombination mechanism (Eq.~\ref{eq:Lrecloc}) accounts for the
electronic excitation decay to the ground state due to nonradiative
processes, and is determined by the recombination time $\tau_{\mathrm{rec}}$,
which is assumed the same for both chromophores. 
The trapping process (Eq.~\ref{eq:LRCloc}) dictates the 
electronic excitation harvesting at the reaction center from the acceptor 
chromophore on a time scale 
determined by the trapping time $\tau_{\mathrm{\mathrm{trap}}}$.
Recombination and trapping times are expected to be on the order of
nanoseconds and picoseconds, respectively \cite{MR&08,ND&10,MKT14}.

\textbf{\textit{Energy transport}}.\textemdash
We assume that only the donor 
chromophore $D$ interacts with the incoherent light and the acceptor 
chromophore $A$ has a lower electronic 
excitation energy, i.e., $\varepsilon_{D}>\varepsilon_{A}$. 
To analyze the transport between chromophores, note that the Hamiltonian 
of the vibronic dimer (Eq.~\ref{eq:Hs}) in the electronic-vibrational 
localized state basis (single-excitation manifold) 
$\{|g_{j}^{\beta}\varepsilon_{i}^{\alpha}\rangle=
|\chi_{j,\beta}^{(g)},g_{j},
\chi_{i,\alpha}^{(\varepsilon)},\varepsilon_{i}\rangle\}$ 
is given by 
\begin{alignat}{1}
\label{eq:H_evloc}
\hat{H}_{\mathrm{Sys}}= & 
\sum_{i\neq j}^{A,D}\sum_{\alpha,\beta}\varepsilon_{i}
|g_{j}^{\beta}\varepsilon_{i}^{\alpha}\rangle\langle\varepsilon_{i}^{\alpha}g_{j}^{\beta}|
+\sum_{i\neq j}^{A,D}\sum_{\alpha,\beta}V_{ij}
|g_{j}^{\beta}\varepsilon_{i}^{\alpha}\rangle\langle\varepsilon_{j}^{\alpha}g_{i}^{\beta}|
\nonumber \\
& +\sum_{i\neq j}^{A,D}\sum_{\alpha,\nu,\mu}\hbar\mathcal{G}_{i}
\left(\delta_{\nu,\mu+1}\sqrt{\mu+1}+\delta_{\nu,\mu-1}\sqrt{\mu}\right)
|g_{j}^{\alpha}
\varepsilon_{i}^{\nu}\rangle\langle\varepsilon_{i}^{\mu}g_{j}^{\alpha}|
+\sum_{i\neq j}^{A,D}\sum_{\alpha,\beta}\alpha\,\hbar\Omega_{i}
|g_{j}^{\beta}\varepsilon_{i}^{\alpha}\rangle\langle\varepsilon_{i}^{\alpha}g_{j}^{\beta}|.
\end{alignat}
The first term of the right-hand site of Eq.~\ref{eq:H_evloc} accounts
for the electronic site energies, the second stands for the excitonic
coupling, the third is the vibronic coupling, and the last is the
vibrational energy.

The change in the population of the state $|g_{D}^{\mu}\varepsilon_{A}^{\nu}\rangle$
(see Eqs. \ref{eq:Mastereq} and \ref{eq:H_evloc})
is equal to 
\begin{alignat}{1}
\label{eq:pop_donorlevel}
\langle g_{D}^{\mu}\varepsilon_{A}^{\nu}|\dot{\hat{\rho}}|
g_{D}^{\mu}\varepsilon_{A}^{\nu}\rangle= & \,
2V_{AD}\Im\left[\rho_{DA}^{\nu,\mu;\nu,\mu}\right]
+ 2\hbar\mathcal{G}_{A}\left(\sqrt{\nu}
    \Im\left[\rho_{AA}^{\nu-1,\mu;\nu,\mu}\right]
+\sqrt{\nu+1}\Im\left[\rho_{AA}^{\nu+1,\mu;\nu,\mu}\right]\right) 
\nonumber \\ &
-\tau_{\mathrm{rec}}^{-1}\rho_{AA}^{\nu,\mu;\nu,\mu}
-\tau_{\mathrm{trap}}^{-1}\rho_{AA}^{\nu,\mu;\nu,\mu}.
\end{alignat}
The unitary contributions to the change in the population of
the acceptor state $|g_{D}^{\mu}\varepsilon_{A}^{\nu}\rangle$ originate from 
the electronic interaction ($V_{AD}$) with the donor state 
$|g_{A}^{\mu}\varepsilon_{D}^{\nu}\rangle$
through the imaginary part of the coherence density matrix element
$\rho_{DA}^{\nu,\mu;\nu,\mu}$, and from the vibronic interaction
$\mathcal{G}_{A}$ with acceptor states with adjacent vibrational
numbers $\nu-1$ and $\nu+1$. 
The non-unitary contributions arise from recombination and trapping. 
It should be noted that the phonon baths do not contribute directly to 
the change of the population of the state 
$|g_{D}^{\mu}\varepsilon_{A}^{\nu}\rangle$ 
due to the pure dephasing interaction in the system-bath Hamiltonian 
\cite{JB20,RW16} (see Eq.~\ref{eq:Hsb}).

By tracing over all the vibrational
states, the change in the population at the acceptor chromophore reads
\begin{align}
\label{eq:pop_site}
\frac{\dd P_{A}}{\dd t} =
\sum_{\nu,\mu}\langle g_{D}^{\mu}\varepsilon_{A}^{\nu}|\dot{\hat{\rho}}|
g_{D}^{\mu}\varepsilon_{A}^{\nu}\rangle =
2V_{AD}\Im\left[\rho_{DA}\right] -(\tau_{\mathrm{rec}}^{-1}+\tau_{\mathrm{trap}}^{-1})\rho_{AA},
\end{align}
where $\Im\left[\rho_{DA}\right]=\sum_{\nu,\mu}\Im
\left[\rho_{DA}^{\nu,\mu;\nu,\mu}\right]$ and 
$\rho_{AA} = \sum_{\nu,\mu}\rho_{AA}^{\nu,\mu;\nu,\mu}$,
which are the intersite coherence and acceptor population of the electronic 
subsystem.
The second term on the right-hand side of Eq.~\ref{eq:pop_donorlevel}
disappears since contributions from different vibrational levels on the
acceptor add up to zero. 
From Eq.~\ref{eq:pop_site} it can be inferred that the flux from the donor 
chromophore is
$F_{DA}=2V_{AD}\Im\left[\rho_{DA}\right]$.
The above allows us to connect with the case of transport
in an electronic dimer without intramolecular vibrational modes \cite{JB20}, i.e. without the effects of vibronic resonance.

\textbf{\textit{Quantum yield: Energy transfer efficiency}.}\textemdash 
We consider a definition for the quantum yield $\eta$ based on 
currents\cite{CB20,JM20a} (transfer rates)
\begin{equation}
\label{eq:QY_def}
\eta=\frac{\mathcal{J}_{\mathrm{trap}}}{\mathcal{J}_{\mathrm{abs}}},
\end{equation}
where the $\mathcal{J}_{\mathrm{trap}}=
\sum_{i}\langle\psi_{i}^{(e)}|
\mathscr{L}_{\mathrm{\mathrm{trap}}}\left[\hat{\rho}\right]
|\psi_{i}^{(e)}\rangle$
is the current of trapping at the reaction center, and 
$\mathcal{J}_{\mathrm{abs}}=
\sum_{i}\langle\psi_{i}^{(e)}|
\mathscr{L}_{\mathrm{\mathrm{RB}}}^{(abs)}\left[\hat{\rho}\right]
|\psi_{i}^{(e)}\rangle$
is the incoherent light absorption current. Note that $\mathcal{J}_{\mathrm{trap}}\neq\tau_{\mathrm{\mathrm{trap}}}^{-1}$
(see Eq.~\ref{eq:LRCloc}).
The states ${|\psi_{i}^{(e)}\rangle}$ are (single-excitation
manifold) eigenstates of the vibronic dimer Hamiltonian, 
$\hat{H}_{\mathrm{Sys}} |\psi_{i}^{(e)}\rangle = E_i |\psi_{i}^{(e)}\rangle$.
While both currents $\mathcal{J}_{\mathrm{trap}}$ and 
$\mathcal{J}_{\mathrm{abs}}$ contribute to the determination of the quantum 
yield, we find that $\mathcal{J}_{\mathrm{abs}}$ is essentially constant.
This suggests that the changes in the quantum yield are mainly dependent on the 
trapping current $\mathcal{J_{\mathrm{trap}}}$.

In the steady-state regime, the following continuity equation must be 
satisfied
\begin{equation}
\label{eq:continuity_eq}
\frac{\mathcal{J}_{\mathrm{emi}} + \mathcal{J}_{\mathrm{rec}}
+\mathcal{J}_{\mathrm{trap}}}{\mathcal{J}_{\mathrm{abs}}}=1.
\end{equation}
Here, $\mathcal{J}_{\mathrm{emi}}=
\sum_{i}\langle\psi_{i}^{(e)}|
\mathscr{L}_{\mathrm{\mathrm{RB}}}^{(emi)}\left[\hat{\rho}\right]
|\psi_{i}^{(e)}\rangle$
and 
$\mathcal{J}_{\mathrm{rec}}=
\sum_{i}\langle\psi_{i}^{(e)}|
\mathscr{L}_{\mathrm{\mathrm{rec}}}\left[\hat{\rho}\right]
|\psi_{i}^{(e)}\rangle$
denote the incoherent light emission and recombination currents, respectively. 
The incoherent light currents $\mathcal{J}_{\mathrm{abs}}$ and 
$\mathcal{J}_{\mathrm{emi}}$ are related to the blackbody radiation bath 
occupation number $(\mathrm{e}^{\hbar\omega/k_{\mathrm{B}}T^{\mathrm{RB}}}
-1)^{-1},\:\omega\geq0$ (absorption) and 
$(\mathrm{e}^{\hbar\omega/k_{\mathrm{B}}T^{\mathrm{RB}}}-1)^{-1}+1,\:\omega<0$
(emission) \cite{CB20}. 
From Eq.~\ref{eq:continuity_eq}, it can be concluded that the quantum yield 
$\eta=1$ when $\mathcal{J}_{\mathrm{emi}}=\mathcal{J}_{\mathrm{rec}}=0$. 
Although, the current definitions discussed above have been described
for the single-excitation eigenstate manifold, their definitions are
basis-invariant \cite{JM20a}, i.e., the sum 
$\sum_{i}\langle\psi_{i}^{(e)}|\dots|\psi_{i}^{(e)}\rangle$
represents the trace over the system of interest.

\textbf{\textit{Quantum yield under an electronic-vibrational resonance 
interaction.}}\textemdash 
\begin{figure}[t]
\includegraphics[scale=1.0]{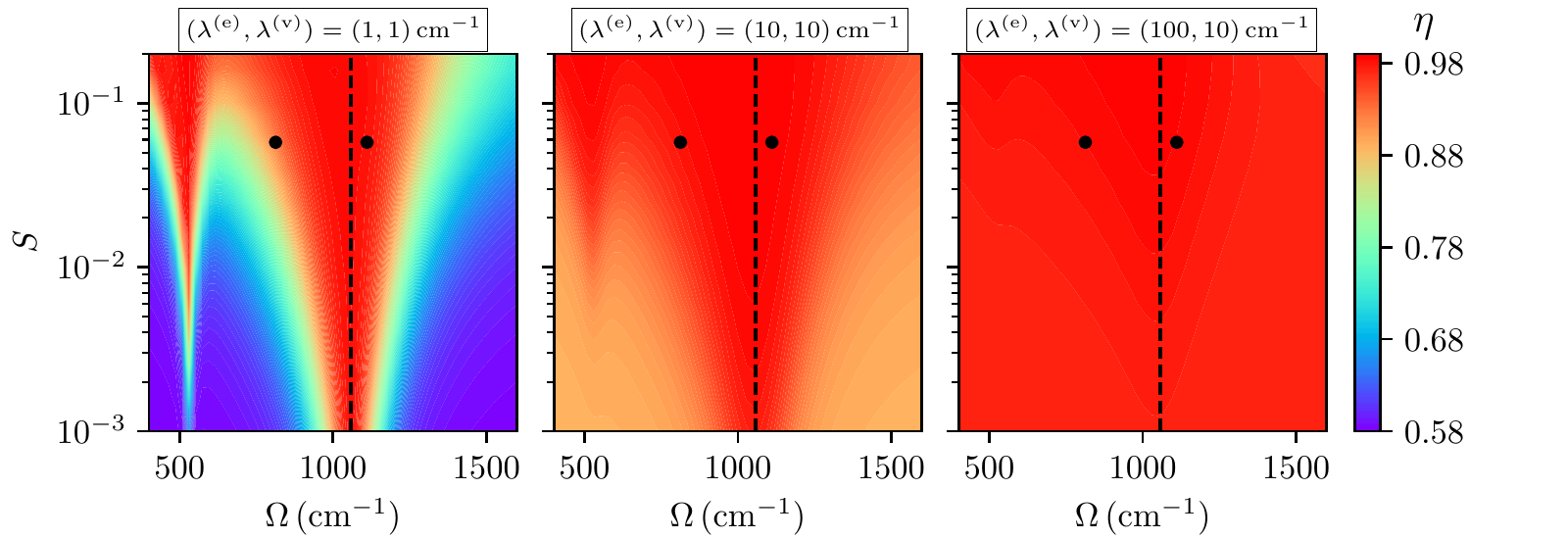} 
\includegraphics[scale=1.0]{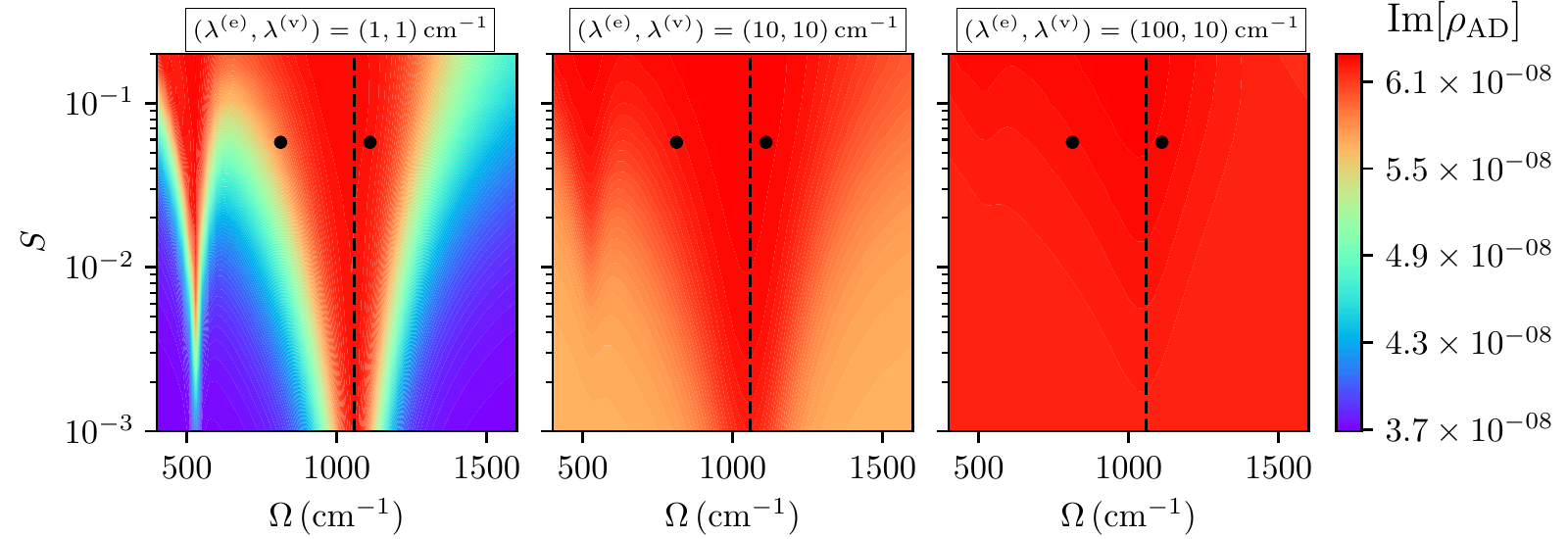} 
\caption{
Quantum yield (top panels) and the imaginary part of the intersite coherence 
(bottom panels) as a function of the Huang-Rhys factor $S$ and the 
intramolecular vibrational frequency $\Omega$ considering different 
reorganization energies $\lambda^{(\mathrm{e},\mathrm{v})}$ 
(system–phonon bath coupling strength). 
Typical values for recombination $\tau_{\mathrm{rec}}=1\,\mathrm{ns}$ and 
trapping $\tau_{\mathrm{trap}}=10\,\mathrm{ps}$ times are assumed.
}
\label{fig:eta_imags12}
\end{figure}
Given the recent interest in the potential quantum effects and energy transfer
enhancements driven by the electronic-vibrational resonance interaction
\cite{YH&19,CC&20,DW&15}, we first compute the steady-state 
$\frac{\partial\rho_{\mathrm{SS}}}{\partial t}=0$
in Eq.~\ref{eq:Mastereq} for different values of the intramolecular
vibrational frequency $\Omega$ close to the exciton 
energy gap $\Delta E_{e}=1058\;\mathrm{cm}^{-1}$ 
(vibronic resonance) and Huang-Rhys factors $S$ (vibronic coupling strengths) 
reported for cryptophyte algae \cite{DM&04,ND&10}.
To quantify a possible effect due to the vibronic resonance, 
Figure~\ref{fig:eta_imags12} shows the quantum yield (top panels) as a 
function of the intramolecular vibrational frequency $\Omega$ and the 
Huang-Rhys factor $S$ for selected values of the system-phonon bath 
coupling strength $\lambda^{(\mathrm{e},\mathrm{v})}$.
We assume typical values for recombination $\tau_{\mathrm{rec}}=1\,\mathrm{ns}$ 
and trapping $\tau_{\mathrm{trap}}=10\,\mathrm{ps}$ times 
(See Figure~S1 presented in the Supporting Information where a trapping 
time $\tau_{\mathrm{trap}}=1\,\mathrm{ps}$ is considered).
A vertical dashed line in Figure~\ref{fig:eta_imags12} indicates the 
on-resonance intramolecular vibrational frequency 
$\Omega = 1058\,\mathrm{cm}^{-1}$, and the two black points mark 
the frequencies $\Omega = 813\,\mathrm{cm}^{-1}$ (off-resonance) and 
$\Omega = 1111\,\mathrm{cm}^{-1}$ (quasi-resonance) for the Huang-Rhys 
factor $S=0.0578$ reported in previous works \cite{ND&10,KO&12}.

In the case of very weak coupling to the phonon baths 
$(\lambda^{(\mathrm{e})},\lambda^{(\mathrm{v})})=(1,1)\,\mathrm{cm}^{-1}$, 
Figure~\ref{fig:eta_imags12} shows a significant 
change in the quantum 
yield and the imaginary part of the intersite coherence in the region close 
to the resonance frequency $\Omega = 1058\,\mathrm{cm}^{-1}$ and for a half 
of the resonance frequency $\Omega = 529\,\mathrm{cm}^{-1}$ for different 
values of the Huang-Rhys factor.
The formation of a resonance pattern is observed in this case. Hence, the 
NESS allows for the vibronic resonance effect.
However, these values for the reorganization energies are far smaller 
than those estimated for the real biophysical systems, 
which are in an intermediate coupling regime with the phonon bath, i.e., 
$\lambda^{(\mathrm{e})}\sim V_{AD}\sim \mathcal{G}_{A/D}$.

\begin{figure}[t]
\includegraphics[scale=1.0]{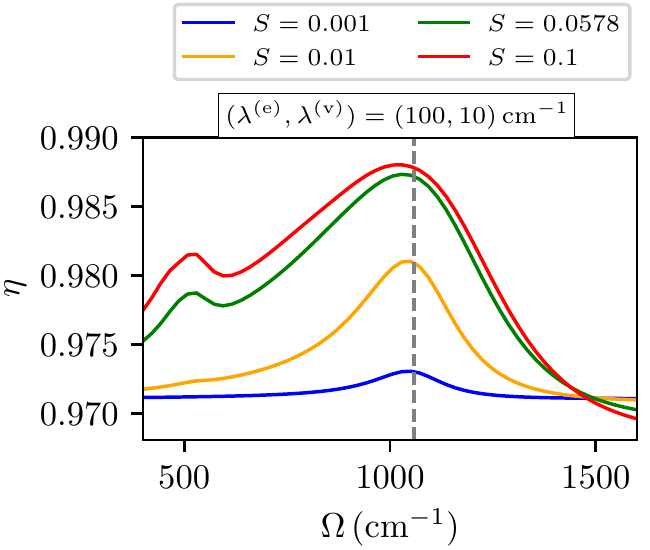} 
\includegraphics[scale=1.0]{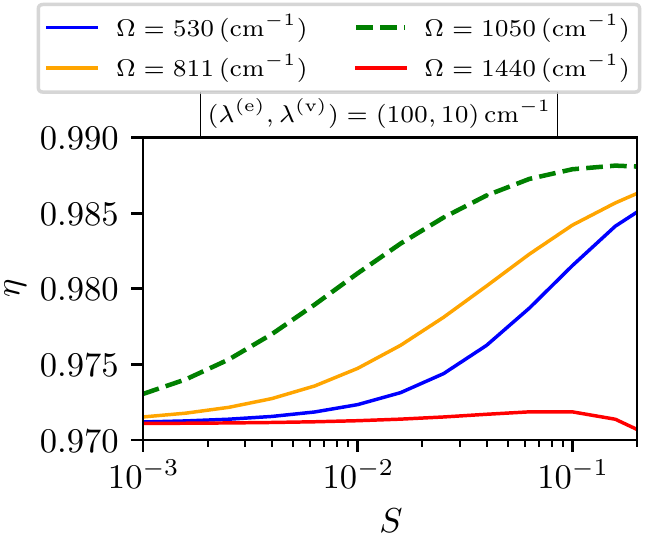}
\caption{
Quantum yield as a function of the intramolecular vibrational 
frequency $\Omega$ (left) and the Huang-Rhys factor $S$ (right) for the case 
$\lambda^{(\mathrm{e})}=100\,\mathrm{cm}^{-1}$ and 
$\lambda^{(\mathrm{v})}=10\,\mathrm{cm}^{-1}$ 
(see the top right panel in Fig. \ref{fig:eta_imags12}). 
Color coding is shown on the top of each figure. 
$\tau_{\mathrm{rec}}=1\,\mathrm{ns}$ and $\tau_{\mathrm{trap}}=10\,\mathrm{ps}$.
}
\label{fig:eta_L100_omega_S}
\end{figure}

\sloppy As can be seen in Figure~\ref{fig:eta_imags12}, as the values 
of system-phonon bath couplings increase, the pattern of the 
electronic-vibrational resonance observed for 
$(\lambda^{(\mathrm{e})},\lambda^{(\mathrm{v})})=(1,1)\,\mathrm{cm}^{-1}$
fades away. 
In fact, for realistic values of the reorganization energies, e.g.,  
$(\lambda^{(\mathrm{e})},\lambda^{(\mathrm{v})})=(100,10)\,\mathrm{cm}^{-1}$, 
as expected for photosynthetic complexes such as the protein-antenna 
phycoerythrin 545,
the changes in the quantum yield due to the vibronic resonance are 
insignificant (see also Figure~\ref{fig:eta_L100_omega_S}).
Therefore, in the case of realistic coupling strengths to the phonon 
environment in the NESS the vibronic resonance effects disappear.
Comparing the upper and the bottom panels in Figure~\ref{fig:eta_imags12} it 
can be concluded that the changes in the quantum yield have the same trend 
as the imaginary part of the intersite coherence, as pointed out 
above in the analysis of energy transport. 
In other words, the current from the donor 
chromophore is proportional to the imaginary part of the intersite coherence 
$F_{DA}=2V_{AD}\Im\left[\rho_{DA}\right]$.

\begin{figure}[t]
\includegraphics[scale=1.0]{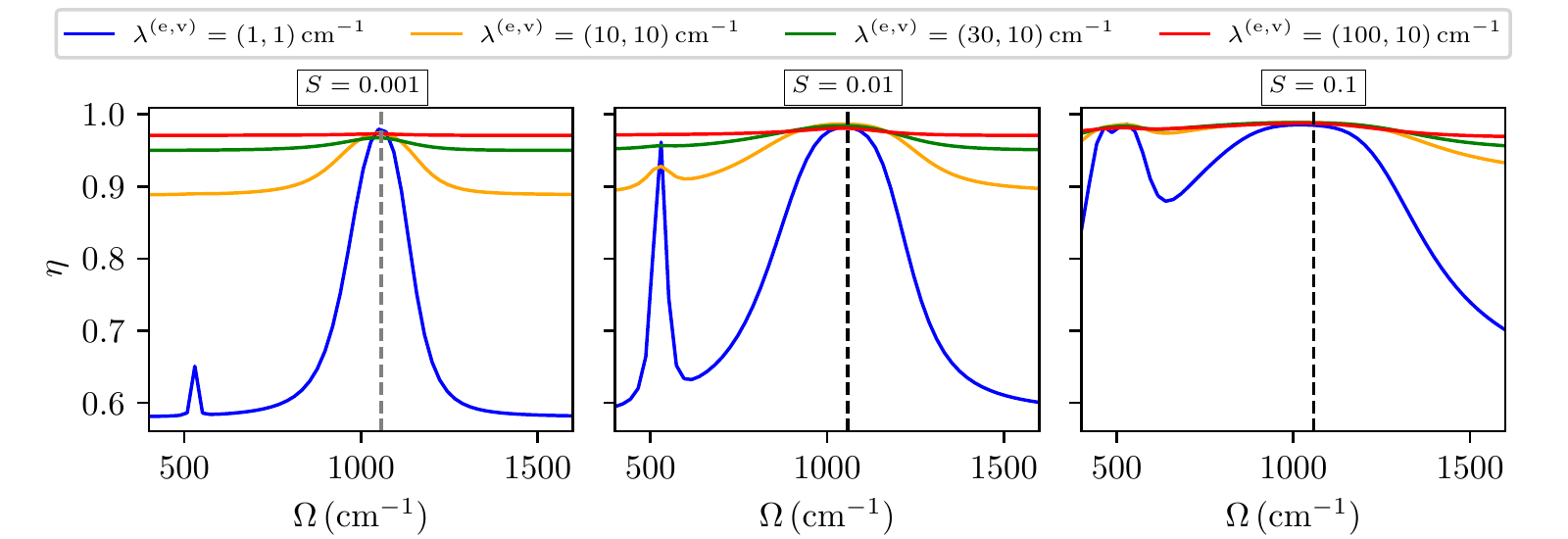} 
\caption{
Quantum yield as function of the intramolecular vibrational frequency 
for specific values of the Huang-Rhys factor $S=0.001$ (left) $S=0.01$ (middle) 
$S=0.1$ (right) and different values of the reorganization energies 
$\lambda^{(\mathrm{e},\mathrm{v})}$ (color coding is shown on the top). 
$\tau_{\mathrm{rec}}=1\,\mathrm{ns}$ and 
$\tau_{\mathrm{trap}}=10\,\mathrm{ps}$.
}
\label{fig:eta_S_lambda}
\end{figure}
This physical picture is corroborated in Figure~\ref{fig:eta_S_lambda},  
which displays the quantum yield $\eta$ as a function of intramolecular 
vibrational frequency $\Omega$ for the Huang-Rhys factors 
$S=\{0.001,\, 0.01,\, 0.1\}$ and different values of the reorganization 
energies (color coded). 
For all the Huang-Rhys factors analyzed, vibronic resonance 
signatures (peaks) lead to significant changes in the quantum yield 
only for weak system-phonon bath couplings.
For the expected physical scenario where
$(\lambda^{(\mathrm{e})},\lambda^{(\mathrm{v})})=(100,10)\,\mathrm{cm}^{-1}$ 
(red lines), there is no effect on the quantum yield, neither with frequency 
changes nor changes in the Huang-Rhys factor.

Figure~\ref{fig:eta_L100_omega_S} shows detailed information on the quantum 
yield as a function of the intramolecular vibrational frequency (left panel) 
and the Huang-Rhys factor (right panel) for the case 
$(\lambda^{(\mathrm{e})},\lambda^{(\mathrm{v})})=(100,10)\,\mathrm{cm}^{-1}$ 
(corresponding to the top right panel in Fig.~\ref{fig:eta_imags12} and red 
lines in Fig.~\ref{fig:eta_S_lambda}). 
Even though small spikes appear close to the resonance region, the changes 
in the quantum yield are marginal for changes in 
the intramolecular vibrational frequency or the Huang-Rhys factor 
(note the vertical scale).
Analogous results are obtained for different recombination and 
trapping times where quantum yields are far less than one.
\emph{Therefore, when all non-unitary contributions in 
Eq.~\ref{eq:Mastereq} are considered, i.e., radiation and phonon baths, 
exciton recombination and trapping at the reaction center, and the physical 
parameters are within the regime expected for the natural function, there is 
no significant changes in the quantum yield resulting from 
vibronic resonance in the NESS} (We present additional NESS results 
supporting the above by using the hierarchical equations of motion (HEOM) 
method in Figure~S2 of the Supporting Information). 

\textbf{\textit{Non-unitary contributions in the establishment of the 
NESS.}}\textemdash 
Below we examine individual terms (recombination, trapping and phonon bath
coupling) in an effort to identify the particular feature (or features) 
that allow the electronic-vibrational resonance effect.
Although the natural scenario corresponds to the
situation discussed above, where the exciton population
is harvested at the reaction center, to quantify a possible effect 
due to the vibronic resonance in the case of no trapping at the reaction 
center, we consider the population in the acceptor chromophore 
$\rho_{AA}$.

\begin{figure}[t!]
\includegraphics[scale=1.0]{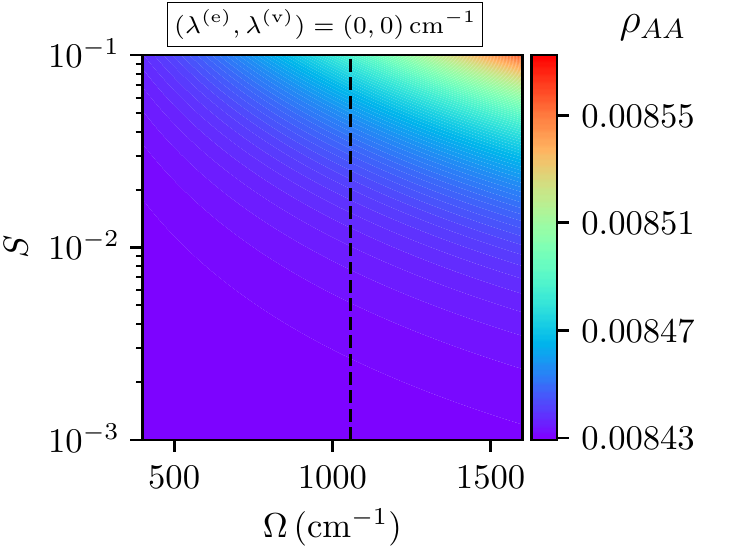}
\includegraphics[scale=1.0]{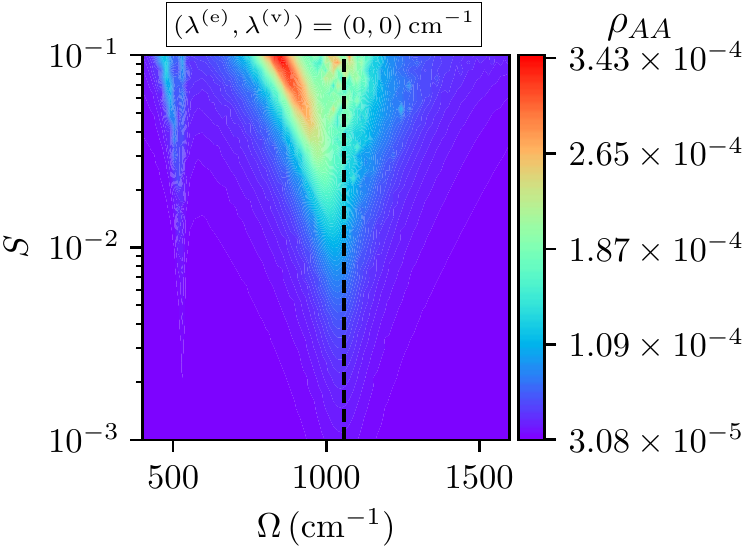} 
\caption{
NESS population of the acceptor chromophore as a function of the 
intramolecular vibrational frequency $\Omega$ and the Huang-Rhys factor $S$
when the coupling strengths to the phonon baths are zero ($\lambda^{(\mathrm{e})}=0$ and 
$\lambda^{(\mathrm{v})}=0$), and there is no trapping at the reaction center.
Left panel: $\tau_{\mathrm{rec}}=\infty$ (Incoherent light only). 
Right panel $\tau_{\mathrm{rec}}=1\,\mathrm{ns}$ (Incoherent light + 
recombination only).
}
\label{fig:s1_onlylight_rec}
\end{figure}
First, consider the case where the system only interacts with the incoherent 
light and reaches a canonical thermal equilibrium at long times
$\hat{\rho}^{(\mathrm{eq,\,RB})} = 
\mathrm{e^{-\beta^{\mathrm{RB}}\hat{H}_{\mathrm{Sys}}}}/
\mathrm{Tr\,\mathrm{e^{-\beta^{\mathrm{RB}}\hat{H}_{\mathrm{Sys}}}}}$,
where $\beta^{\mathrm{RB}}=1/(k_{\mathrm{B}}T^{\mathrm{RB}})$. 
Due to the high temperature of the blackbody radiation bath 
$T^{\mathrm{RB}}=5600\,\mathrm{K}$,
higher energy eigenstates get a larger population (e.g., compared
with the case when the phonon baths are turned on 
($T^{\mathrm{PB}}=300\,\mathrm{K}$).
Thus, as expected for the vibronic dimer examined here, 
Figure~\ref{fig:s1_onlylight_rec} (left panel) shows that the population of 
the acceptor chromophore increases for higher intramolecular vibrational 
frequencies and Huang-Rhys factors. 
However, the population of the acceptor chromophore does not peak when 
the frequency of the intramolecular vibrational mode is in resonance with 
the exciton energy gap, as seen in the left panel of 
Figure~\ref{fig:s1_onlylight_rec}.
That is, the distinctive resonance pattern observed in 
Figure~\ref{fig:eta_imags12} for the case of very low reorganization energies 
is nonexistent when the system thermalizes with the incoherent photon bath 
only or, more generally, any singular thermal bath. 
Hence, the equilibrium state reached under the effect of the incoherent 
radiation bath is not sensitive to the vibronic resonance effect.

However, if the system interacts with the incoherent light and is also
subjected to recombination, the corresponding NESS allows for the vibronic 
resonance effect by facilitating the transport to the ground 
vibrational level in the electronically excited state of the acceptor 
chromophore, as seen in the right panel of Figure~\ref{fig:s1_onlylight_rec}.
That is, the population of the ground vibrational level in the electronically 
excited state of the acceptor increases in the resonance, leading to a deviation 
from the thermal population $\hat{\rho}^{(\mathrm{eq,\,RB})}$.
Meanwhile, the acceptor population increases when the energy frequency of the 
intramolecular vibrational mode is in resonance with the exciton energy gap, 
indicating that there is a vibronic resonance effect.
Furthermore, the population in the acceptor chromophore is larger with 
increasing Huang-Rhys factor. 
On the other hand, for very low Huang-Rhys factors ($S\leq0.001$), the 
vibronic resonance effect is barely noticeable since the vibronic couplings 
are too small. 
Similar observations can be made if the system interacts with the incoherent 
light and is subjected to recombination and trapping 
(see Figure~S3 in the Supporting Information). 
Formally, when $\tau_{\mathrm{rec}}\to\infty$ and
$\tau_{\mathrm{trap}}\to\infty$, then
$\mathcal{J}_{\mathrm{rec}}\to0$ and
$\mathcal{J}_{\mathrm{trap}}\to0$;
the NESS corresponds to an equilibrium thermal
state $\rho_{\mathrm{SS}}\to\ensuremath{\rho^{(\mathrm{eq,\,RB})}}$,
and therefore $\mathrm{Im}[\rho_{AD}]\to0$. 
Therefore, \emph{when the reached NESS is close to an equilibrium state, the 
signature pattern of the electronic-vibrational resonance does not arise.}

\sloppy When the effects of incoherent light, recombination,
and phonon baths are considered, there is no change in the 
acceptor NESS population associated with the electronic-vibrational resonance 
interaction. 
Specifically, \emph{it is the coupling to the phonon baths that suppresses 
vibronic resonance effects for realistic reorganization energies} (e.g.,
$(\lambda^{(\mathrm{e})},\lambda^{(\mathrm{v})})=(100,10)\,\mathrm{cm}^{-1}$).
In particular, the interaction with the phonon baths is
stronger than the interaction with incoherent light \cite{PBB17,PB13}, so that 
the resonance-induced effects observed in the case of 
`incoherent light + recombination only' owing to the vibronic coupling 
are overshadowed by the dissipative effect of the intermolecular vibrational 
degrees of freedom associated with the low-frequency phonon baths.

\textbf{\textit{Time-dependent results: Coherent initial excitation.}}\textemdash 
\begin{figure}[t]
\includegraphics[scale=1.]{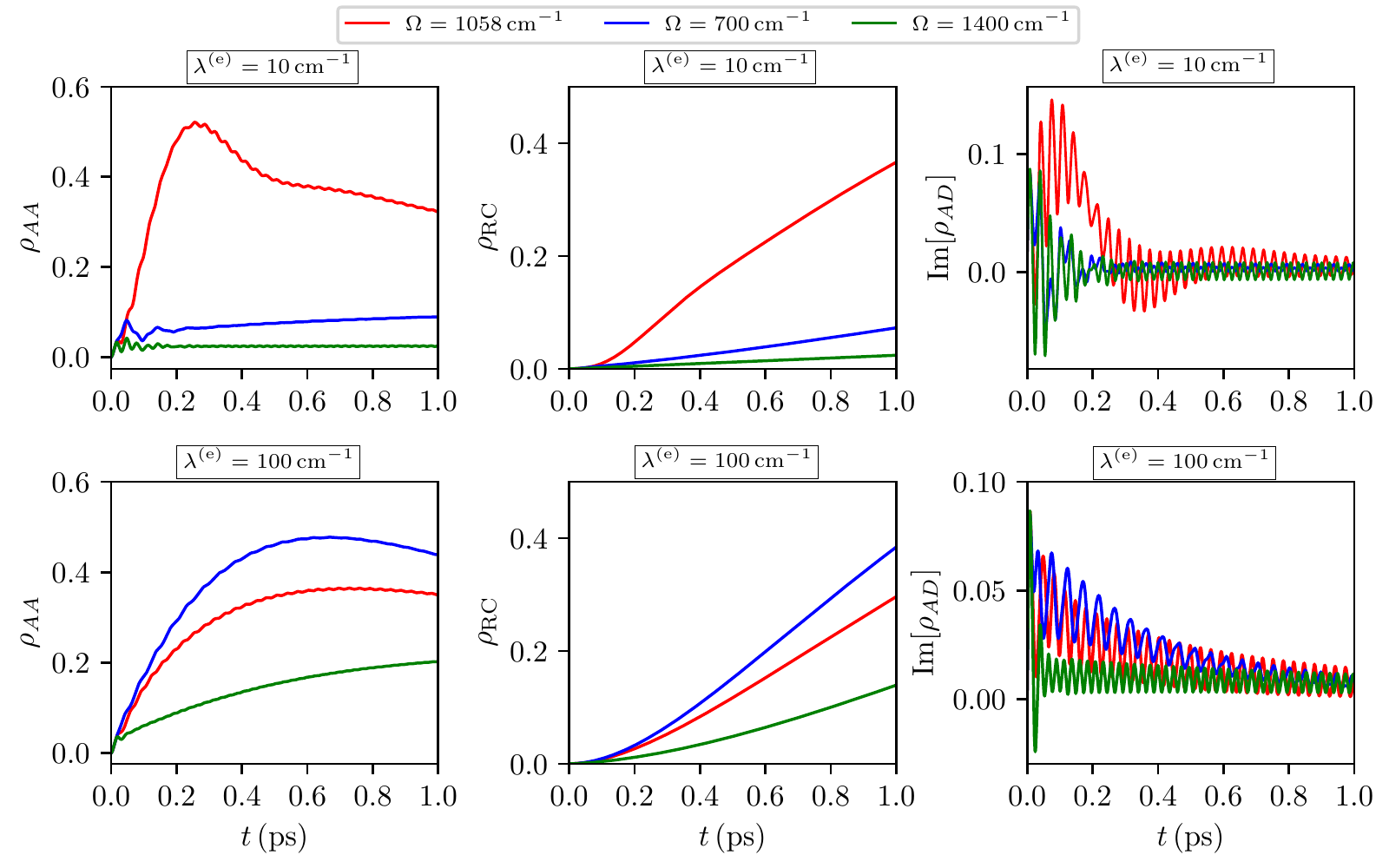}
\caption{
Population of the acceptor chromophore, population of the reaction center, 
and the imaginary part of the intersite coherence as a function of time 
for different values of the intramolecular vibrational 
frequency $\Omega$ (note the color coding). 
Two scenarios for the system-phonon bath coupling are considered: small  
$\lambda^{(\mathrm{e})}=10\,\mathrm{cm}^{-1}$ (top panels) and intermediate 
$\lambda^{(\mathrm{e})}=100\,\mathrm{cm}^{-1}$ (bottom panels).
For all cases, the vibronic coupling is the same 
$\mathcal{G}=250\,\mathrm{cm}^{-1}$.
$\tau_{\mathrm{rec}}=1\,\mathrm{ns}$ and 
$\tau_{\mathrm{trap}}=1\,\mathrm{ps}$.
}
\label{fig:heom_dynamics}
\end{figure}
To assess the similarities regarding the vibronic resonance between  
NESS results discussed above
and time-dependent results related to coherent light excitation discussed in 
the literature, Figure~\ref{fig:heom_dynamics} shows the population 
(acceptor and reaction center) and intersite coherence (imaginary part) 
dynamics for times up to one picosecond, assuming an initial coherent excitation 
condition $\rho_{D}(t=0)=1$, i.e., the entire population initially 
in the donor.
The top and bottom panels in Figure~\ref{fig:heom_dynamics} display the dynamics 
using the HEOM method (see Supporting Information)
for weak and intermediate system-phonon bath coupling 
$\lambda^{(\mathrm{e})}=10\,\mathrm{cm}^{-1}$ 
($\lambda^{(\mathrm{e})}=100\,\mathrm{cm}^{-1}$) for three 
intramolecular vibrational frequencies: $\Omega=1058\,\mathrm{cm}^{-1}$ 
(on-resonance) and $\Omega=\{700,\,1400\}\,\mathrm{cm}^{-1}$ (off-resonance).
The same vibronic coupling $\mathcal{G}=250\,\mathrm{cm}^{-1}$ is considered 
in all the cases, and typical values for recombination 
$\tau_{\mathrm{rec}}=1\,\mathrm{ns}$ and trapping 
$\tau_{\mathrm{trap}}=1\,\mathrm{ps}$ times are assumed.

For weak system-phonon bath coupling, the time-dependent results 
evidence an enhancement in the acceptor and the reaction center population, as 
well as in the imaginary part of the intersite coherence related to the flux 
between chromophores for the on-resonance vibrational 
frequency compared to the off-resonance frequencies 
(see top panels in Figure~\ref{fig:heom_dynamics}).
However, for intermediate system-phonon bath coupling, the increase 
in the acceptor and the reaction center population, as 
well as in the imaginary part of the intersite coherence, is 
not guaranteed for the vibrational frequency on-resonance with the 
exciton energy gap 
(see red curves in bottom panels of Figure~\ref{fig:heom_dynamics}).
In fact, higher values in the acceptor and the reaction center population, 
as well as in $\Im\left[\rho_{DA}\right]$, are obtained for the 
off-resonance frequency 
$\Omega=700\,\mathrm{cm}^{-1}$ (see blue curves in bottom panels of 
Figure~\ref{fig:heom_dynamics}). 
This is in contrast to prior work \cite{OO14} which did not examine the lower
off-resonance frequencies (here $\Omega=700\,\mathrm{cm}^{-1}$) and concluded 
that a significant resonance effect exists.

We observe then that for weak system-phonon bath coupling the time-dependent 
results evidence an electronic-vibrational resonance effect, as 
observed in the case of the NESS results discussed above.
For realistic values of the reorganization energy, that is, in the 
intermediate system-phonon bath coupling regime, the results differ from the NESS
case, in that time-dependent results could lead to significantly different 
population and coherence dynamics as a function of the intramolecular 
vibrational frequency.
However, it is possible to obtain higher values for the 
acceptor and the reaction center population for off-resonance vibrational 
frequencies as compared to the on-resonance vibrational frequency.
These observations merit further study, ongoing in our laboratory.

\textbf{\textit{Conclusions.}}\textemdash 
Natural incoherent light-induced processes in photosynthetic 
light-harvesting systems occur in a non-equilibrium steady-state (NESS). 
Here, the extent to which electronic-vibrational resonance associated with 
high-energy intramolecular vibrations affects the steady-state energy 
transport in the PEB dimer was analyzed using a prototype vibronic dimer, 
i.e., two two-level systems, each coupled to an intramolecular 
vibrational mode. 
\textit{The results indicate that in the NESS for the PEB dimer, 
intramolecular vibrations with frequency resonant with the energy difference 
between exciton states do not alter the quantum yield nor 
the imaginary part of intersite coherence relevant for transport compared 
to non-resonant vibrations.}
The results clearly motivate studies of other relevant 
biological models.

Individual non-unitary contributions (incoherent light, 
recombination, trapping and phonon coupling) were analyzed to determine the 
physical feature that destroys the electronic-vibrational resonance effect.
We conclude that in the natural scenario of incoherent light excitation  
and under realistic system-phonon bath couplings, the steady-state reached 
is dominated mainly by the effect of the phonon baths, and the impact of 
the vibronic resonance is not evident.
Moreover, under initial coherent excitation, as
prepared by pulsed lasers and commonly assumed in other studies about
vibronic effects in photosynthetic light-harvesting systems, and considering 
an intermediate system-phonon bath coupling regime, it is possible 
to obtain higher values for the acceptor population and the 
imaginary part of the intersite coherence
for off-resonance vibrational frequencies as 
compared to the on-resonance vibrational frequency.
These PEB results imply that energy transport is not
dramatically affected by electronic-vibrational 
resonance.

\begin{acknowledgement}
This work was supported by the U.S. Air Force Oﬃce of Scientific  
Research (AFOSR) under Contract No. FA9550-20-1-0354.
\end{acknowledgement}

\begin{suppinfo}
\label{suppinfo}
Electronic dimer model; Quantum yield analysis considering a trapping time 
of one picosecond; presentation of NESS results using the HEOM method; 
Figure S3: quantum yield for the case of incoherent light + recombination 
+ trapping only.
\end{suppinfo}

\bibliography{refs}

\end{document}